\documentclass[aps,prl,preprint,nopacs,superscriptaddress]{revtex4}

\usepackage{graphicx}
\usepackage{verbatim}
\usepackage{mathrsfs}
\usepackage{gensymb}
\usepackage{color}
\usepackage[colorlinks,bookmarks=false,citecolor=blue,linkcolor=blue,urlcolor=blue]{hyperref}
\usepackage{amsmath,amsfonts,amssymb}

\newcommand{\bee}{\begin{equation}}
\newcommand{\ee}{\end{equation}}
\def\3{2.8in}    
\def\2{2.5in}
\def\4{3.0in}\def \beq {\begin{equation}}
\def \eeq {\end{equation}}
\begin{document}

\title{Tunable disorder and localization in the rare-earth nickelates}
\author{Changan Wang}\email{changan.wang@hzdr.de}
\affiliation{Helmholtz-Zentrum-Dresden-Rossendorf, Bautzner Landstra{\ss}e 400, 01328 Dresden, Germany}
\affiliation{Technische Universit\"at Dresden, 01062 Dresden, Germany}
\affiliation{Shenzhen Key Laboratory of Laser Engineering, College of Optoelectronic Engineering, Shenzhen University, 518060 Shenzhen, China}
\author{Ching-Hao Chang}\email{c.h.chang@ifw-dresden.de}
\affiliation{Leibniz-Institute for Solid State and Materials Research, Helmholtzstra{\ss}e 20, 01069 Dresden, Germany}
\author{Angus Huang}
\affiliation{Department of Physics, National Tsing Hua University, Hsinchu 30043, Taiwan}
\author{Pei-Chun Wang}
\affiliation{Department of Materials Science and Engineering, National Chiao Tung University, Hsinchu, Taiwan}
\author{Ping-Chun Wu}
\affiliation{Department of Materials Science and Engineering, National Chiao Tung University, Hsinchu, Taiwan}
\author{Lin Yang}
\affiliation{Institute for Advanced Materials and Guangdong Provincial Key Laboratory of Quantum Engineering and Quantum Materials, South China Normal University, Guangzhou 51006, China}
\author{Chi Xu}
\affiliation{Helmholtz-Zentrum-Dresden-Rossendorf, Bautzner Landstra{\ss}e 400, 01328 Dresden, Germany}
\affiliation{Technische Universit\"at Dresden, 01062 Dresden, Germany}
\author{Parul Pandey}
\affiliation{Helmholtz-Zentrum-Dresden-Rossendorf, Bautzner Landstra{\ss}e 400, 01328 Dresden, Germany}
\author{Min Zeng}
\affiliation{Institute for Advanced Materials and Guangdong Provincial Key Laboratory of Quantum Engineering and Quantum Materials, South China Normal University, Guangzhou 51006, China}
\author{Roman B\"{o}ttger}
\affiliation{Helmholtz-Zentrum-Dresden-Rossendorf, Bautzner Landstra{\ss}e 400, 01328 Dresden, Germany}
\author{Horng-Tay Jeng}
\affiliation{Department of Physics, National Tsing Hua University, Hsinchu 30043, Taiwan}
\affiliation{Institute of Physics, Academia Sinica, Taipei 11529, Taiwan}
\author{Yu-Jia Zeng}
\affiliation{Shenzhen Key Laboratory of Laser Engineering, College of Optoelectronic Engineering, Shenzhen University, 518060 Shenzhen, China}
\author{Manfred Helm}
\affiliation{Helmholtz-Zentrum-Dresden-Rossendorf, Bautzner
Landstra{\ss}e 400, 01328 Dresden, Germany}
\affiliation{Technische Universit\"at Dresden, 01062 Dresden,
Germany}
\author{Ying-Hao Chu}
\affiliation{Department of Materials Science and Engineering, National Chiao Tung University, Hsinchu, Taiwan}
\affiliation{Center for Emergent Functional Matter Science, National Chiao Tung University, Hsinchu 30010, Taiwan}
\author{R. Ganesh}
\affiliation{The Institute of Mathematical Sciences, HBNI, C I T Campus, Chennai 600113, India}
\author{Shengqiang Zhou}
\affiliation{Helmholtz-Zentrum-Dresden-Rossendorf, Bautzner Landstra{\ss}e 400, 01328 Dresden, Germany}

\date{\today}

\begin{abstract}
The rare-earth nickelates are a rich playground for transport
properties, known to host non-Fermi liquid character, resistance
saturation and metal-insulator transitions. We report a study of transport in LaNiO$_3$ in the presence of tunable disorder induced by irradiation.
While pristine LaNiO$_3$ samples are metallic, highly irradiated samples
show insulating behaviour at all temperatures. Using irradiation
fluence as a tuning handle, we uncover an intermediate region
hosting a metal-insulator transition.
This transition falls within the Mott-Ioffe-Regel regime wherein the mean free path is comparable to lattice spacing. In the high temperature metallic regime, we find a transition from non-Fermi liquid to a Fermi-liquid-like character. On the insulating side of the metal-insulator transition, we find behaviour that is consistent with weak localization. This is reflected in magnetoresistance that scales with the square of the field and in resistivity. In the highly irradiated insulating samples, we find good agreement with variable range hopping, consistent with Anderson localization. We find qualitatively similar behaviour in thick PrNiO$_3$ films as well. Our results demonstrate that ion irradiation can be used to tailor transport, serving as an excellent tool to study the physics of localization.
\end{abstract}
\maketitle 

\paragraph{Introduction}
A central question in condensed matter physics is the distinction
between insulators and metals. In this context, Anderson
localization \cite{and58, lee85, Kramer1993} is a key paradigm
that explains how a disordered potential can suppress transport
via quantum interference. This is particularly interesting in
three spatial dimensions wherein a threshold disorder strength is
required to localize electrons. Direct simulations of 3D Anderson
localization  have been achieved in areas as diverse as light
\cite{Wiersma1997}, sound \cite{Hu2008} and ultracold atom gases
\cite{Kondov66,Jendrzejewski2012}. Signatures of localization have
also been seen in solid state systems, mainly in doped
semiconductors \cite{Katsumoto1987,Waffenschmidt1999,Itoh2004,Mobius2018}. In
this letter, we demonstrate that irradiation can serve as a tuning
knob to study a wide range of localization phenomena.

We study transport in the rare-earth nickelates
\cite{Medarde1997}. This family of materials hosts several complex
phenomena such as non-Fermi liquid character \cite{jaramillo2014},
charge fluctuations \cite{Johnston2014}, magnetic correlations
\cite{Guo2018} and a width-tuned MIT (metal-insulator transition)
\cite{sch11}. As the underlying physics continues to be debated,
we show that these materials can be tuned across a wide range of
localization phenomena by irradiation. 
We explain our results using several known paradigms: the Mott-Ioffe-Regel limit, weak localization and variable range hopping.


Since the formulation of the iconic Drude model, it is known that
a key quantity that determines conductivity is mean-free path --
the average distance travelled by electrons between successive
collisions. The collisions themselves can be elastic (between
electrons and crystal defects) or inelastic (electron-electron or
electron-phonon scattering). We use ion irradiation to induce
defects \cite{gia11,guo15}, thereby directly controlling the
elastic scattering rate. Our results demonstrate that this
provides a superior tuning knob for transport properties when
compared to tuning strain, doping or thickness. Unlike these
methods, irradiation tunes the mean free path without
significantly affecting other system parameters such as
dimensionality and carrier concentration.


\paragraph{Experimental details}
LaNiO$_3$ (LNO) films of width 50 nm were epitaxially grown on
SrTiO$_3$ (001) single crystals ($a_{sub} = 3.905~\AA$) by pulsed
laser deposition. The films were then irradiated by 6 keV He ions
with four different fluences, $\Phi = 1$, $1.75$, $2.5$, to
$5\times10^{15}$ He/cm$^{2}$. Our work is, in part, motivated by
early studies on ternary borides where He-irradiation was shown to
affect transport \cite{Rowell1980}. The He ions create point
defects with a density that increases with fluence. Helium's
nobility ensures that no extra holes or electrons are doped into
the films. It does induce strain that is relieved by an increase
of out-of-plane lattice constant (see Supplementary
Figure 1), but this doesn't significantly affect carrier density.
The pristine and irradiated samples were characterized by x-ray
diffraction (XRD) (PANalytical XÕPert PRO diffractometer) using Cu
K$\alpha$ radiation. Transport properties were measured using a
Van-der-Paul geometry and a constant current source. To measure
temperature dependence, temperature was swept at a slow rate of
1-2 K/min with a Lakeshore 332 temperature controller. We have
performed the same measurements on PrNiO$_3$ as well, with
qualitatively similar results.

\begin{figure}
\includegraphics[width=5.5 in]{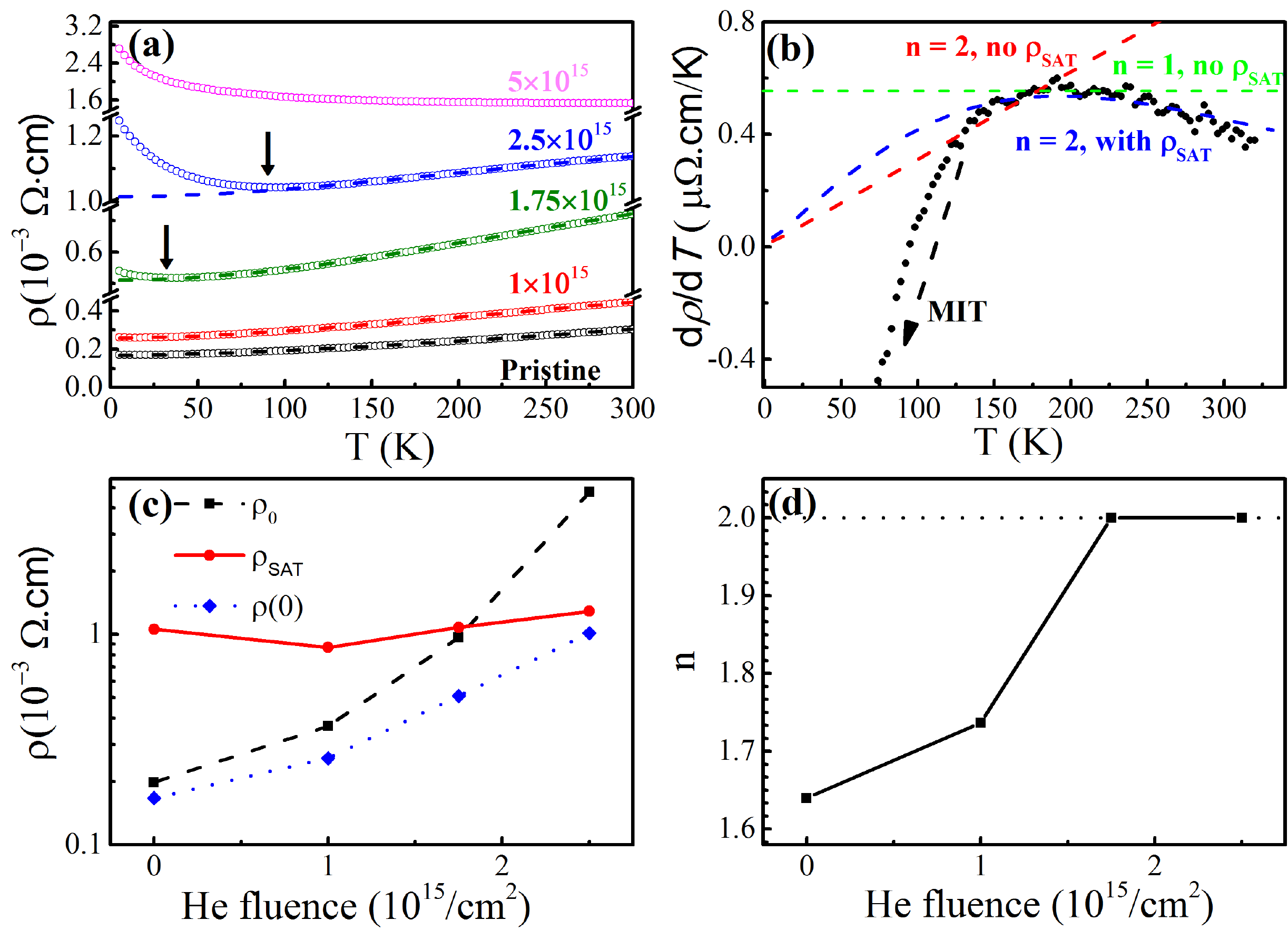}
\caption{(a) Resistivity vs. temperatures of LNO samples with the increasing fluence of irradiation. Arrows mark MITs. The dashed lines are fits of data in the metallic region using Eq. (\ref{eq.rhoideal}) and Eq. (\ref{eq.rhocomb}). (b) Temperature derivative of the resistivity as a function of temperature for the sample with a fluence $\Phi = 2.5 \times10^{15}$ He/cm$^{2}$. 
The downturn at lower temperature represents the MIT. Red and green dashed lines are fits with $\rho_{\rm sat}=0$, which do not describe the data well.
(c) $\rho_0$,  $\rho_{\rm sat}$, and $\rho(T = 0)$ obtained at different fluences. (d) Extracted exponents $n$ for different fluences. }
\label{fig1}
\end{figure}

\paragraph{Resistivity}
Metallic and insulating behaviours are distinguished by the
temperature dependence of resistivity. Fig.~\ref{fig1}(a) shows
resistivity vs. temperature for the pristine film as well as
irradiated samples. The pristine film as well that with the lowest
fluence are both metallic at all temperatures. In contrast, the
sample with the largest fluence ($\Phi = 5\times10^{15}$
He/cm$^{2}$) exhibits insulating behavior at all temperatures. At
intermediate fluences, we see a clear minimum in the resistivity,
indicating an MIT. Here, we use the term MIT to denote a thermal crossover 
where the the slope of resistivity vs. temperature changes sign. 
We observe that the MIT temperature increases with increasing fluence.

\paragraph{Metallic side of the MIT}
In a typical metal, resistivity increases with temperature according to
\begin{equation}
\rho_{ideal} (T) = \rho_0+A T^n.
\label{eq.rhoideal}
\end{equation}
The resistivity at zero temperature, $\rho_0$, arises due to
\textit{elastic} scattering from defects in the crystal -- a
measure of disorder strength. In contrast, the temperature
dependent contribution arises largely from \textit{inelastic}
electron-electron and electron-phonon scattering \cite{giu82}. In
a Fermi liquid, when electron-electron scattering is dominant, the
exponent is $n=2$ leading to a characteristic quadratic behaviour.
The interplay of elastic and inelastic scattering can be described
in terms of the mean free path, $\ell$, with $\rho \propto
\ell^{-1}$ \cite{bee91},
\begin{eqnarray}
\frac{1}{\ell} = \frac{1}{\ell_{e}} + \frac{1}{\ell_{in}},
\end{eqnarray}
as the elastic ($\ell_{e}$) and inelastic ($\ell_{in}$) scattering
rates combine additively. With increasing temperature, $\ell_{in}$
decreases leading to increasing resistivity. This increase cannot
go on indefinitely as the mean free path cannot be shorter than
the lattice spacing. The regime when the mean free path is
comparable to the lattice spacing is called the Mott-Ioffe-Regel
limit \cite{gun03,hus04}. This leads to an upper bound on the
resistivity, $\rho_{sat}$, which can be thought of arising from
Heisenberg's uncertainty principle \cite{wie77,gur81,gun03}. The
overall resistivity is given by
\begin{eqnarray}
\rho^{-1}(T) =\rho^{-1}_{ideal}(T)+\rho^{-1}_{\rm sat}.
\label{eq.rhocomb}
\end{eqnarray}
This parallel-resistor formula is well known to describe a wide
variety of metals \cite{gun03}.

Here, we find that nickelate films on the metallic side of the MIT are well described by a combination of Eqs.~\ref{eq.rhocomb} and \ref{eq.rhoideal}.
We fit the resistivity data taking 
$\rho_0$, $\rho_{sat}$, $A$ and $n$ as fitting parameters. The obtained results are shown in Fig.~\ref{fig1}(c,d).
In order to obtain a reasonable fit, it is essential to include a saturation resistivity $\rho_{sat}$ using the parallel resistor formula of Eq.~\ref{eq.rhocomb} \cite{mik15}. This is brought out in Fig.~\ref{fig1}(b) which shows the derivative of resistivity and the corresponding fitting curves. 
As shown in Fig.~\ref{fig1}(c), the resistivity extrapolated to zero temperature, $\rho(0)$, exhibits the typical result of parallel resistor addition. For low fluences, when $\rho_0 \ll \rho_{sat}$, $\rho(0)$ tracks $\rho_0$. At high fluences, when $\rho_0 \gg \rho_{sat}$, $\rho(0)$ approaches $\rho_{sat}$.

As shown in Fig.~\ref{fig1}(c), $\rho_0$ increases with fluence -- consistent with our expectation that $\rho_0$ is determined by the density of defects which controls the elastic mean free path. As $\rho_0$ also depends on carrier density, we measured Hall resistivity on the pristine, $\Phi = 1\times10^{15}$ and $\Phi = 1.75\times10^{15}$ He/cm$^{2}$ samples at 75 K (all three samples are metallic at this temperature). The densities inferred from the Drude formula are 2.02, 1.40 and 1.31 ($\times 10^{22} \mathrm{cm}^{-3}$) respectively. This shows that changes in transport are driven by changing defect density, rather than carrier density. For example, the $\Phi = 1\times10^{15}$ (always metallic) and $\Phi = 1.75\times10^{15}$ (hosting an MIT) samples only differ by $\sim 7\%$ in carrier density, but by $\sim 60\%$ in $\rho_0$.

The saturation resistivity $\rho_{sat}$ is more or less
independent of fluence, in line with our expectation that
$\rho_{sat}$ is an intrinsic material property that depends on the
lattice constant (and perhaps the carrier density). Earlier
studies have suggested that $\rho_{sat}$ in nickelate films is
sensitive to the in-plane strain \cite{mik15} as it modifies bands
near the Fermi energy \cite{Yoo2015}. In our samples, the in-plane
strain is fixed by the STO substrate and does not vary with
irradiation (see Supplementary Figure 2). Indeed, our
$\rho_{sat}$ value is close to that seen in thick RNO films
\cite{mik15}.

The fluence-dependence of $\rho_0$ and $\rho_{sat}$ indicates
progression towards the Mott-Ioffe-Regel limit and beyond. When
$\rho_0$ exceeds $\rho_{sat}$ at a fluence of $\Phi \approx
1.75\times10^{15}$ He/cm$^{2}$, it suggests that the elastic mean
free path becomes comparable to the lattice spacing (we support
this assertion with further arguments below).
Surprisingly, this is reflected in the exponent $n$, plotted in Fig.~\ref{fig1}(d). For low fluences, the exponent is $n\approx 1.6$ indicating non-Fermi liquid (NFL) behaviour, known to be present in LNO. This could possible arise from lattice \cite{jaramillo2014}, charge \cite{Johnston2014} or magnetic fluctuations \cite{Guo2018,Subedi2017}. 
At large fluences with $\Phi \gtrsim 1.75 \times10^{15}$
He/cm$^{2}$, we find $n\approx 2$ suggesting Fermi liquid (FL)
character. Remarkably, this NFL-FL transition coincides with the
Mott-Ioffe-Regel (MIR) limit as suggested by $\rho_0$ and
$\rho_{sat}$ values. This suggests that a large disorder
concentration is required to suppress magnetic/lattice
fluctuations. NFL character in the nickelates and in
heterostructures has seen a surge of interest with many recent
experimental studies \cite{mik15,Zhang2016,Phanindra2018}. Our
results suggest that this behaviour can be tuned by disorder
concentration and indeed, vanishes in the MIR limit.

\begin{figure}
\includegraphics[width=5.5in]{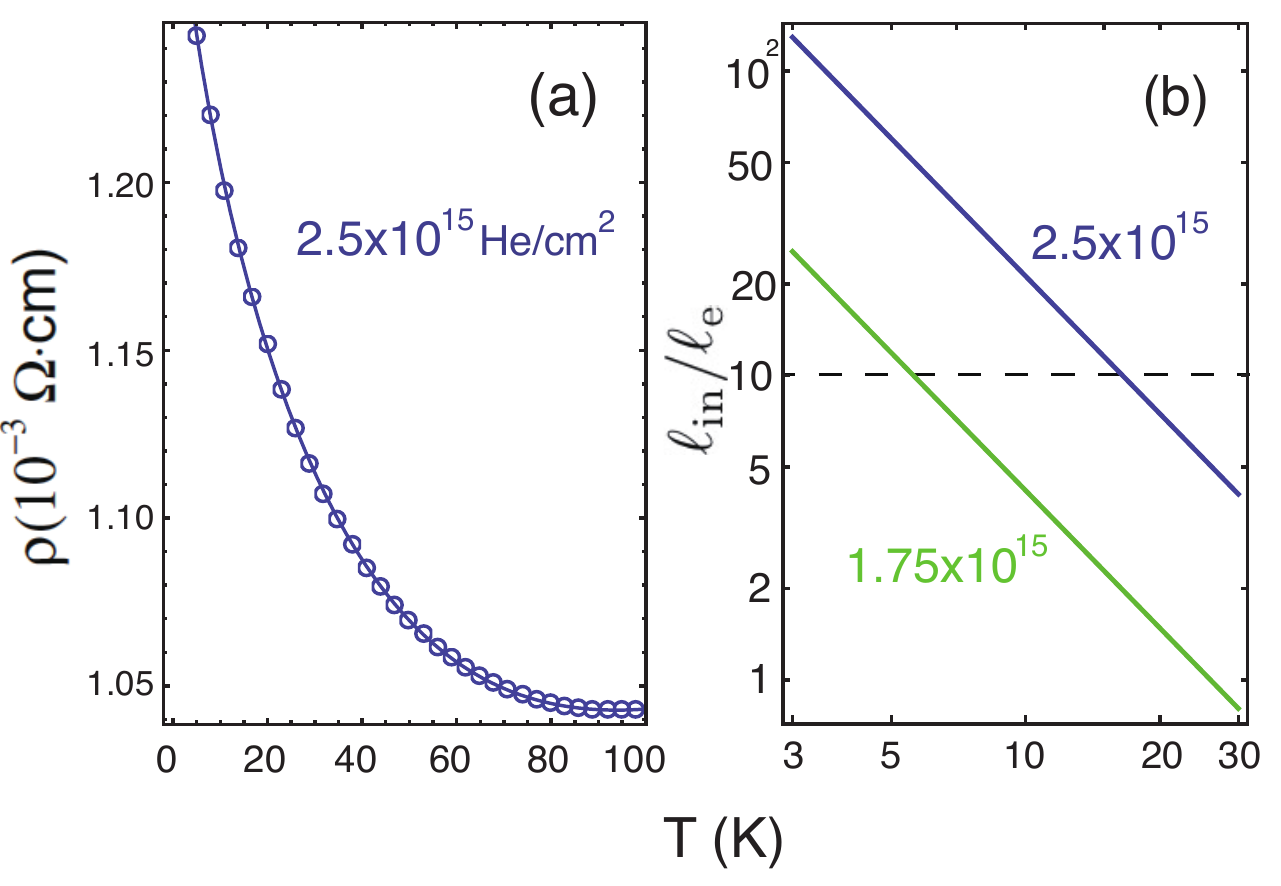}
\caption{(a) Increase of resistivity on the insulating side of the MIT, with a fit to Eq.~\ref{eq:weak}.
(2) A log-log plot of the ratio of mean free paths, obtained from the fitting parameters using Eq.~\ref{eq:lr}.}
\label{fig2}
\end{figure}

\paragraph{Insulating side of the MIT} Having discussed the MIT seen at intermediate fluence values ($\Phi = 1.75$ and $ 2.5 \times10^{15}$ He/cm$^{2}$), we characterize the insulating side of the transition. In three dimensions, weak disorder (with strength below the threshold required for Anderson localization) leads to increased resistance due to quantum interference. In this regime, we argue that resistivity can be described by (see Supplementary Section 2)
\begin{align}
\rho (T) = \rho(0)- \alpha T^{n/2}+\beta T^{n}-\gamma T^{3n/2},
\label{eq:weak}
\end{align}
where $\alpha$, $\beta$, and $\gamma$ are coefficients related to the ratio of inelastic and elastic mean free paths, $\ell_{\rm in}/\ell_{\rm e}$,
\begin{align}
\frac{\ell_{\rm in}(T)}{\ell_{\rm e}} = \frac{1}{\pi}\left( \frac{\beta^2}{\alpha^2}  -\frac{\gamma}{\alpha}\right)^{-1}T^{-n}.
\label{eq:lr}
\end{align}
In Fig.~\ref{fig2}(a), the resistivity is shown with the best fit
to Eq. (\ref{eq:weak}). The best fit coefficient here is $n =
1.5$, indicating that the inelastic scattering process is
dominated by electron-electron interactions ($\ell_{\rm in}
\propto T^{-2}$) \cite{giu82,bab37} rather than electron-phonon
interactions ($\ell_{\rm in} \propto T^{-3}$) \cite{ram86}. Our
result is consistent with previous observations in LNO films
\cite{sch11,wei15}. The ratio $\ell_{\rm in}/ \ell_{\rm e}$ from
Eq. (\ref{eq:lr}) obtained from best fit parameters is shown in
Fig.~\ref{fig2}(b). The increase of He fluence from  $\Phi = 1.75$
to $ 2.5 \times10^{15}$ He/cm$^{2}$ increases defect
concentration, thereby decreasing $\ell_e$ and increasing
$\ell_{in}/\ell_{e}$. The obtained ratio $\ell_{\rm in}/\ell_{\rm
e}$ allows us to make quantitative statements about approaching
the MIR limit. For instance, in the sample with $\Phi = 2.5
\times10^{15}$ He/cm$^{2}$ with temperature around $\sim 3 K$, we
find $\ell_{\rm in}/\ell_{\rm e} \gtrsim 100$ as shown in
Fig.~\ref{fig2}(b). In this temperature range, the inelastic mean
free path ($\ell_{\rm in}$) for an LNO film is known to be
approximately 40 nm \cite{sch11,her05}. Therefore, the elastic
mean free path ($\ell_{\rm e}$) is as short as 4 \AA, close to the
LNO lattice constant 3.83 \AA. This confirms that the film is
indeed at the MIR limit. From the large value of $\ell_{\rm
in}/\ell_{\rm e}$, we also deduce that inelastic scattering
processes are not important at low temperatures. It follows that
the resistivity increase (with decreasing temperature) in this
regime stems from elastic scattering and the associated quantum
interference effects. This is consistent with our analysis
assuming weak localization.

 \begin{figure}
\includegraphics[width=3.5in]{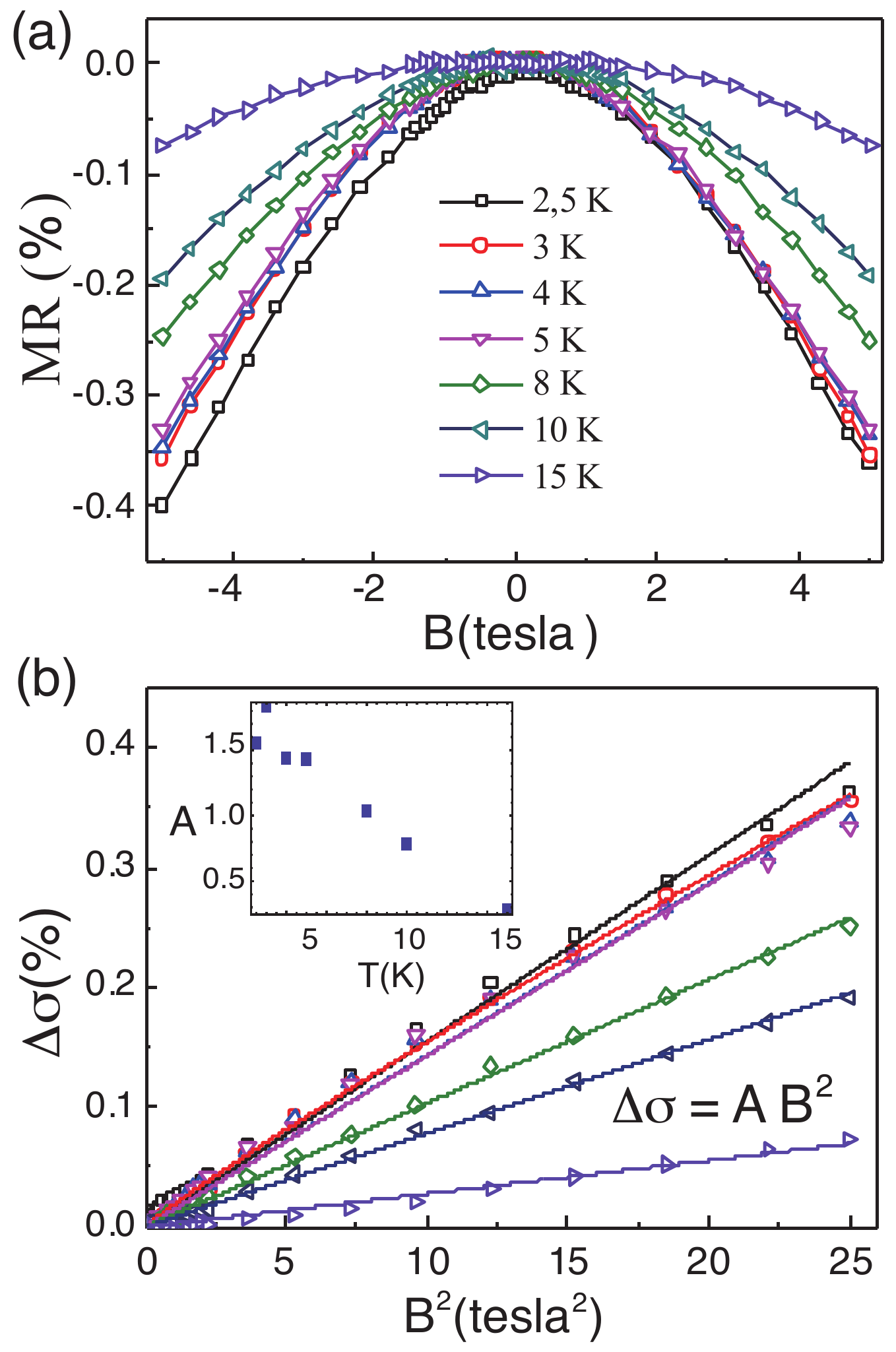}
\caption{(a) Magnetoresistance ($\{R(H)-R(0)\}/R(0)$) vs. field in the LNO sample irradiated with fluence $2.5\times10^{15}$ He/cm$^{2}$ at different temperatures.
(b) Fractional increase in magnetoconductivity $(\Delta\sigma(\%)= \{\sigma(B)-\sigma(0)\}/\sigma(0))$ vs. the square of the magnetic field.
The inset shows the fitting parameter $A$ (unit: 10$^{-4}$/tesla$^2$) as a function of temperature.}
\label{fig3}
\end{figure}

To further characterize the insulating side of the MIT, we present
magnetotransport data. Fig.~\ref{fig3}(a) shows magnetoresistance
(MR) at different temperatures for the sample irradiated with
fluence $\Phi = 2.5\times10^{15}$ He/cm$^{2}$. From the above
discussion, this sample is expected to be close to the MIR limit.
Resistance decreases with the application of a field as
time-reversal-breaking disrupts quantum interference effects that
are responsible for weak localization. In our samples that lie in
the MIR regime, the elastic mean free path is a few Angstroms. The
magnetic length, however, is $\ell_B \approx 26/\sqrt{B ({\rm
tesla})}$ nm which is always much greater than $\ell_e$. In this
regime, with $\ell_B \gg \sqrt{ \ell_{e} \times \ell_{in}}$, the
magnetoconductivity scales as \cite{kaw80}
\begin{align}
\Delta \sigma(\%)= \frac{\sigma(B,T)}{\sigma(0,T)}-1 \approx \Big(\frac{\ell_{\rm in}}{ \ell_{\rm e}}\Big)^{3/2}\Big(\frac{e\tau_{\rm e}}{m^\star}\Big)^2 B^2,
\label{eq:mc}
\end{align}
where $m^\star$ is the electron effective mass and $\tau_{\rm e} =
\ell_{\rm e}/v_F$ is the elastic relaxation time proportional to
the Fermi velocity $v_F$. Notably, the magnetoconductivity scales
as the square of the field. Our data is in good agreement with
this form as shown in Fig.~\ref{fig3}(b); we see that $B^2$
scaling persists over a large field range, up to 5 $T$. Moreover,
the fitting parameter $A$ in Fig.~\ref{fig3}(b) decreases with the
increase of temperature. This is consistent with Eq.~\ref{eq:mc}
as the ratio $\ell_{in}/\ell_{e}$ decreases with temperature (see
Fig.~\ref{fig2}(b)). The magnetoconductivity of the sample
irradiated with flux $\Phi = 1.75\times10^{15}$ He/cm$^{2}$ also
shows $B^2$ scaling, as shown in the Supplementary Figure 3. We
reiterate that the $B^2$ scaling is a consequence of the MIR
limit. In pristine LNO, with a field of several Teslas, the
magnetoconductivity scales as $B^{1/2}$ as the mean free path is
longer than the magnetic length \cite{her05}.

\begin{figure}
\includegraphics[width=3.5in]{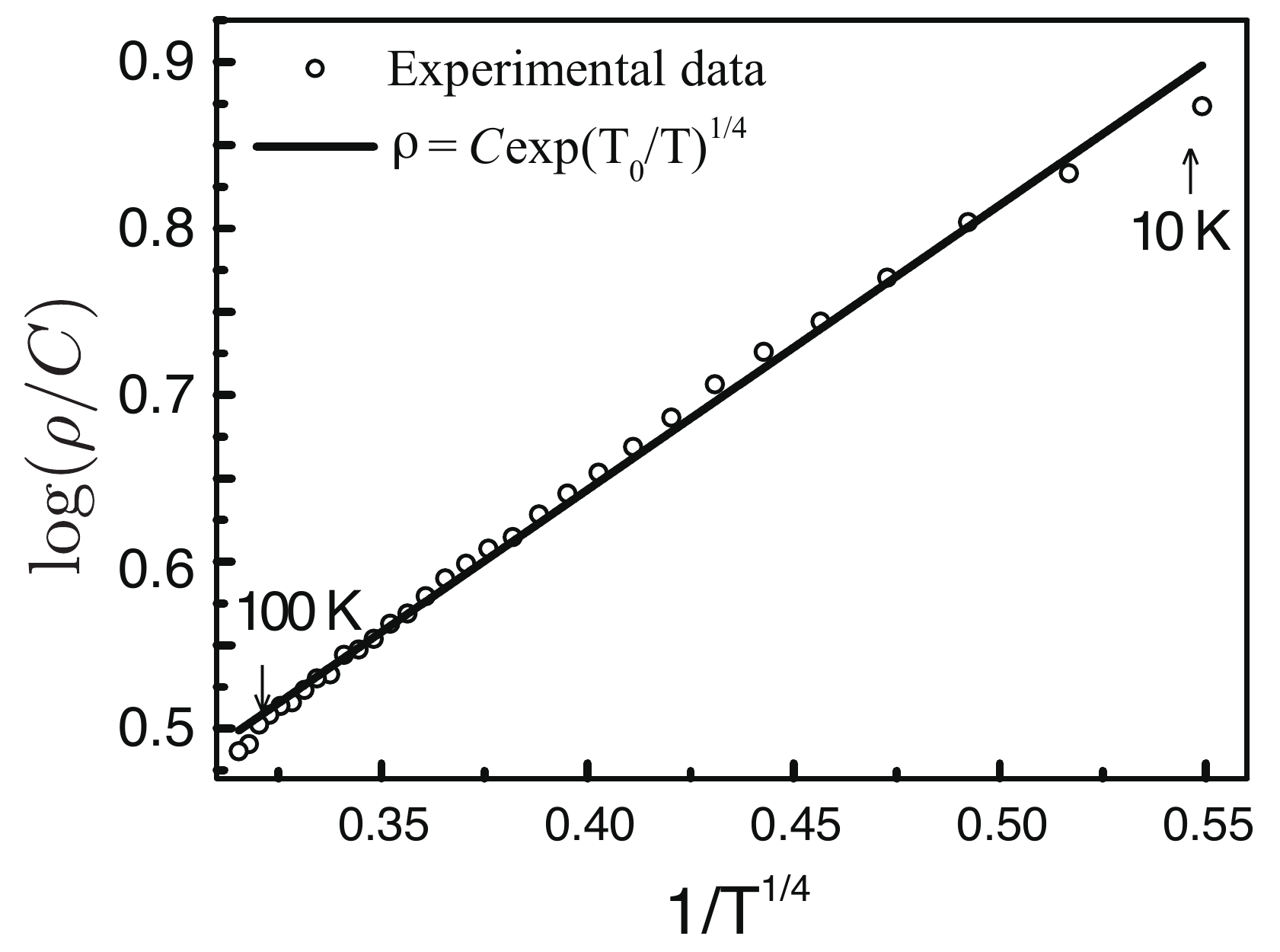}
\caption{Logarithm of resistivity vs. T$^{-1/4}$ for the sample irradiated with fluence $5\times 10^{15}$ He/cm$^{2}$. The black line is a fit to the data between 10 and 100 K.}
\label{fig4}
\end{figure}

\paragraph{Insulating samples} Having discussed both sides of the MIT seen at intermediate fluence, we discuss the sample with highest fluence, $\Phi = 5\times10^{15}$ He/cm$^{2}$ (see pink circles in Fig.~\ref{fig1}(a)). This shows insulating behaviour at all temperatures. With high disorder concentration, we expect electrons to be Anderson-localized. Transport can only occur via hopping between localized states that are well-separated in position and energy, equivalent to a percolation process \cite{aps75}. In this scenario, resistivity follows the variable range hopping (VRH) paradigm \cite{mot69}, with
\begin{align}
\rho(T) = C {\rm exp}(T_0/T)^{1/(1+d)},
\label{eq:insulator}
\end{align}
where $d$ is the number of spatial dimensions. $T_0$ is a constant that depends on both the density of localized states and the spatial decay profiles of wavefunctions. Fig.~\ref{fig4} shows our resistivity data with a fit to Eq.~\ref{eq:insulator} with $d=3$. We obtain a good fit from 10 to 100 K.
Earlier studies on insulating ultrathin LNO films that are
insulating have shown good agreement with VRH behaviour, described
by Eq. (\ref{eq:insulator}) with $d = 2$ \cite{bre73,sch11}. The
two-dimensionality arises from lateral confinement
\cite{ort11,cha17}. Here, as our film thickness is 50 nm, much
greater than the mean free path, we find three-dimensional
character.

\paragraph{Discussion} We have studied transport in nickelate films, with tunable disorder. With irradiation as a tuning knob, we see (i) a metallic phase in the clean regime, (ii) metal-insulator transitions for intermediate disorder, and (iii) insulating behaviour for high disorder. 
Our results on transport in LaNiO$_3$ are robust, with the same qualitative results occurring in irradiated PrNiO3 as well (see Supplementary Figure 4),
and can be compared with earlier studies on the nickelates using various tuning parameters: electric fields \cite{sch09}, oxygen vacancies \cite{jaramillo2014}, strain \cite{mik15} and film thickness \cite{sch11,Pontes2014}. We argue that irradiation provides a superior tuning knob as it changes only the elastic mean free path, without strongly affecting carrier density, dimensionality or band energies.

Our findings suggest that irradiated nickelates open a new window
to the study of Anderson localization, as they allow for a clean
handle on the mean free path. In particular, the following
exciting question arises: can we detect a `mobility edge'
\cite{mott2012,Mott1987} in a solid state system? In our samples
with intermediate disorder, a new temperature scale (associated
with the MIT) emerges. Measurements such as optical conductivity
could reveal whether this is correlated with a mobility edge in
the electronic spectrum. Such measurements are of great interest
in the context of recent ultracold atom experiments
\cite{Semeghini2015,Pasek2017}.

\paragraph{Acknowlegements}
C.W. thanks financial support from the China Scholarship Council
(File No. 201606750007). C.-H.C. and P. P. acknowledge financial
support from the German Research Foundation (Deutsche
Forschungsgemeinschaft, Grant CH 2051/1-1 and ZH 225/6-1.).
C.-H.C., A.H., and H.-T.J. acknowledge support from the National
Center for Theoretical Sciences. R.G. thanks IFW Dresden for
hospitality.
\\
C.W. and C.-H.C. contributed equally to this work.

\paragraph{Author contributions} 
C.W., C.-H.C. and S.Z. conceived the project. M. H. supervised the work. C.W. carried out the transport experiments. P.-C.W., P.-C.W., and Y.-H.C. prepared the samples. L.Y., M.Z. and Y.Z. performed XRD measurements.  R.B. performed ion irradiation. C.W. and C.-H.C. carried out data analysis. C.-H.C. and R.G. developed the theoretical models, while A.H. and H.-T.J. performed band structure calculations. C.-H.C., R.G. and S.Z. wrote the paper. All authors discussed the results and commented on the manuscript.

\paragraph{Competing Interests} 
The authors declare no competing interests.


\end{document}